\documentclass[12pt,aps,preprint,epsfig]{revtex4}
\usepackage{graphicx}
\usepackage{epsfig}
\usepackage{amsmath}

\begin{document}

\title{Effective Attraction Interactions between Like-charge Macroions Bound to Binary Fluid Lipid Membranes}

\author{Xia-qing Shi and Yu-qiang Ma}
\altaffiliation[ ]{Author to whom correspondence should be
addressed. Electronic mail: myqiang@nju.edu.cn.}

\affiliation{National Laboratory of Solid State Microstructures,
Nanjing University, Nanjing 210093, China}

\date{\today}

\begin{abstract}
   Using  integral equation theory of liquids to a
binary mixed fluid lipid membrane, we  study the membrane-mediated
interactions between the macroions and the redistribution of
neutral and charged lipids due to binding macroions. We find that
when the concentration of binding macroions is infinitely dilute,
the main contribution to the attractive potential between
  macroions is the line tension between neutral and charged
lipids of the membrane, and the bridging effect also contributes
to the attraction. As  the relative concentration of   charged
lipids is increased, we observe a repulsive - attractive -
repulsive potential transition due to the competition between the
line tension of lipids and screened electrostatic
macroion-macroion interactions. For the finite concentration of
macroions, the main feature of the attraction is similar to the
infinite dilution case.  However, due to the interplay of
formation of   charged  lipid - macroion complexes, the line
tension of redistributed binary lipids induced
 by single macroion is lowered in this case, and
the maximum of attractive potential will shift toward the higher
values of the charged lipid concentration.


\end{abstract}

\maketitle

\section{Introduction}
Generally, it is believed that, in the physiological condition,
biomembranes are quasi two-dimensional (2D) fluid mixtures,
composed of a wide variety of protein and lipid molecular species.
The protein-protein, protein-lipid, and lipid-lipid interactions
in and on the membranes are very important to the functions of the
cell, such as ligand-receptor interactions (1), membrane rafts
(2), and the formation of lipids domains and membrane budding (3),
etc. Most of these problems are complex and not well understood
yet. Recently, there is an increasing interest in understanding
the electrostatic binding of charged macroions to the oppositely
charged lipid membrane both experimentally and theoretically
(4-13). Because of lipid's fluidity in the membrane, when the
charged macroions are bound to the membrane, the oppositely
charged lipids would migrate to the binding sites. This process
would compete with the mixing entropic effects of different lipid
species, and at equilibrium, the minimum of free energy is
required. There are several theoretical works characterizing such
a process (9-13). One of the models takes into account   the
entropy contributions of lipids and membrane-associated proteins
to the free energy
  in the incompressible limit. Combined with the nonlinear
Poisson-Boltzmann (PB) equation, May and Ben-Shaul have obtained the
local lipid composition and the adsorption free energy in the single
protein case, and accessed the influence on these physical
quantities when the concentration of   binding proteins is finite
(9). By introducing a parameter $\chi$ characterizing the extent of
non-ideal lipid mixing in the mean field approximation, they were
able to give the phase behavior of
 charged lipid membranes with the binding peripheral proteins.
The existence of a two-dimensional phase separation suggests that
there is an effective lateral attractive potential between the
binding proteins. However, the nonlinear PB theory itself does not
introduce such a potential (14), and thus the authors argued that
the attraction must be mediated by the membranes (10,12). To
numerically solve the nonlinear PB equation simply and
efficiently, most of the theoretical work mentioned above adopted
cylindrical symmetry with the axis through the center of the
macroion and normal to the charged membrane. Such a symmetric
consideration make it difficult to calculate the effective pair
potential between the macroions. In fact, there have been no
systematic theoretical studies into detailed calculation of
effective interactions between binding macroions. The specific
form of membrane-mediated like-charge attractive potential has not
been clarified, and the mechanism  is still unclear.

In the present  paper, we establish a simple model of the fluid
membrane and the binding macroions, to examine the lipid-mediated
effective pair potential between  binding macroions. Actually, the
calculation of the effective pair potential is very important to
understanding the phase behavior of soft  matters. Recently, the
effective potential in ionic solution that is beyond the
description of the Derjaguin-Landau-Verwey-Overbeek (DLVO) theory
receives great recognition (15), and it is believed that it will
have further applications in biological systems (16,17). The most
counterintuitive phenomenon is the effective attraction between
like-charge objects. Most of these phenomena can be understood as
the strong correlation effects of   multivalent counter-ions
(18-21). However, the long-range attraction between like-charge
spheres confined in parallel glass walls is still an open question
both for experimental and theoretical studies (22-24). It is
believed that such a long-range attraction is mediated by the
substrate parallel glass walls because of confinement, but the
mechanism is still not revealed theoretically (16). On the
contrary,   the theoretical understanding of effective potential
between the macroions mediated by the fluid membrane will be much
more complicated, but is relatively unexplored. There are several
approaches to get the effective potential theoretically. Here we
will adopt the integral equation theory which is an accurate and
powerful tool to explore the microscopic short-range structure of
liquids (25,26), and recently, it has been applied successfully to
explain the haloing effects of colloidal stabilization (27,28).

\section{Model and Theory}
It is well-known that biomembranes are a kind of lyotropic liquid
crystals, and the lipid is an amphiphilic molecule composed of a
hydrophilic head and one or two hydrophobic tail chains. In our
model, we take the membrane  as two-dimensional fluids composed of
 neutral and  charged  lipid species. The interactions between the lipids can be separated into two
parts which are head-head and tail-tail interactions. The
head-head interactions are treated as 2D hard core, while the
tail-tail interactions are treated as 2D soft core to account for
the repulsions between the tails originated from both enthalpy and
entropy of   chains. Thus  the total pair potential
$V_{\alpha\beta}(r)$ between $\alpha$ and $\beta$ lipids with a
lateral separation $r$ can be written as:
\begin{equation} \label{eq:V1}
V_{\alpha\beta} (r) = \left\{ \begin{array}{ll}
     \infty ,   &r \leq r_{\alpha h}+r_{\beta h},\\
    \epsilon_{\alpha\beta}+v_{\alpha\beta}(r),   & r_{\alpha h}+r_{\beta h}< r \leq r_{\alpha s}+r_{\beta s},\\
      v_{\alpha\beta}(r) ,      &r>r_{\alpha s}+r_{\beta s} ,\\
\end{array} \right.
\end{equation}
where the indices $\alpha$ and $\beta$ ($=c$ or $n$), indicate
charged or neutral lipids.   $r_{\alpha h}$ is $\alpha$ lipid's
hard-core radius, and  $r_{\alpha s}$ is $\alpha$ lipid's
soft-core radius. $\epsilon_{\alpha\beta}$ is the soft repulsion
between $\alpha$ and $\beta$ lipids,  therefore the interaction
parameter characterizing the demixing extent of non-ideal mixed
lipids is given by
$\omega_{cn}=\epsilon_{cn}-(\epsilon_{cc}+\epsilon_{nn})/2$. For
the sake of  simplicity, we set $\epsilon_{cc}=\epsilon_{nn}=0$,
and thus $\omega_{cn}=\epsilon_{cn}$. However, we should point out
that the real interaction between chains may be more complex as
presented in literatures (29-31). $v_{\alpha\beta}(r)$ is the
effective potential between the lipids introduced by charged heads
in the absence of binding macroions, and it is obvious that
$v_{nn}(r)=v_{cn}(r)=0$.

The binding macroions can move, but confined in a two-dimensional
  plane which is parallel to
the membrane but with a separation $h$,  as reported in literature
(13). For simplicity, the macroion is modelled as a hard sphere,
which is uniformly charged on the surface. The pair potential
$V_{pp}(r)$ between the macroions can be written as:
\begin{equation} \label{eq:V2}
V_{pp} (r) = \left\{ \begin{array}{ll}
     \infty ,   &r \leq 2r_{p},\\
    v_{pp}(r),   &r>2 r_{p},\\
\end{array} \right.
\end{equation}
 where $r_p$ is the hard-core radius of the macroion, $v_{pp}(r)$ is the screened electrostatic
 interaction in the absence of  charged lipids. In the binding state, there is
another pair potential to account for the interactions between the
negatively charged lipids and positively charged macroions.
Formally, it can be written as:
\begin{equation}\label{eq:V3}
V_{pc}(r)=v_{pc}(r,h),
\end{equation}
where $r$ is the lateral separation between the lipids and
macroions, and $h$ is the minimal membrane-macroion distance (see
Fig. 1). When $h$ is fixed, the pair potential is only concerned
with lateral distance $r$, so such a system can be effectively
treated as a 2D fluid mixtures with pair potentials given above.
Now, we can apply the two-dimensional integral equation theory to
our model. The Ornstein-Zernike (OZ) equations for homogeneous
mixtures in two-dimensional case are given by
\begin{equation}
h_{\alpha\beta}(r)=c_{\alpha\beta}(r)+\sum_{\nu}\sigma_\nu\int
c_{\alpha\nu}(|\textbf{r}-\textbf{r}'|)h_{\nu\beta}(\textbf{r}')\mathrm{d}\textbf{r}',
\end{equation}
where $h_{\alpha\beta}(r)$ is defined by $h_{\alpha\beta}(r)\equiv
g_{\alpha\beta}(r)-1$. $g_{\alpha\beta}(r)$ is the radial
distribution function, and $h_{\alpha\beta}(r)$ and
$c_{\alpha\beta}(r)$ are the total and direct correlation function
between two particles of species $\alpha$ and $\beta$, respectively.
$\sigma_\nu$ is the number density of species $\nu$. The summation
is over all species, and the integral is performed over the
two-dimensional space. Eq.(4) can be closed by the well-known
Percus-Yevick (PY) or hypernetted-chain (HNC) approximations or
other thermodynamic self-consistent approximations (25,26,32) which
have different bridge functions $B_{\alpha\beta}(r)$. Formally, it
can be written as follows:
\begin{equation}\label{bridge}
g_{\alpha\beta}(r)=e^{-\beta
V_{\alpha\beta}(r)+h_{\alpha\beta}(r)-c_{\alpha\beta}(r)+B_{\alpha\beta}(r)}\;.
\end{equation}
In the HNC approximation, $B_{\alpha\beta}(r)=0$, while in the PY
approximation,
$B_{\alpha\beta}(r)=\ln(1+\gamma_{\alpha\beta}(r))-\gamma_{\alpha\beta}(r)
$  where $\gamma_{\alpha\beta}(r)\equiv
h_{\alpha\beta}(r)-c_{\alpha\beta}(r)$.

\section{Details of Calculation}
 The first step to do our calculation is to write
down the specific form of  pair potentials. Here, however, we must
fall back on some approximations. The most important thing that we
care is whether the charges are screened near the membrane  in the
physiological condition. One might argue that since the charges
cannot penetrate into the membrane,    the electro-static
interactions between charged macroions or lipids near the membrane
will not be screened. Such a statement does not take into account
the specific properties of the membrane and the solution. We know
that the hydrophobic part of the cell membrane has a dielectric
constant $\varepsilon_m\sim 2.1$ and a thickness around $6 \sim 7$
nm, while the solution, in the physiological condition, has a
dielectric constant $\varepsilon_s\sim 80$ and the screening
length $\lambda\sim 1$ nm (33). When a charge $Q$ occurs in the
solution near the membrane (within $3 \sim 4$ nm), it will form an
image charge in the membrane with
$Q'=\dfrac{\varepsilon_s-\varepsilon_m}{\varepsilon_s
+\varepsilon_m}Q\approx Q$. Now most of screening charges are near
the membrane within 1 nm, and these charges would form images in
the hydrophobic part of the membrane. Therefore the charges are
surrounded by   screening charges even on the membrane, and the
electrostatic interactions must be more or less screened. This
justifies the use of  screened potentials. The screened potentials
can be written as follows:
\begin{equation} \label{eq:model}
 \left\{ \begin{array}{ll}
     \beta v_{pp}(r)=\dfrac{Z_p^2L_B}{r}e^{-r/\lambda},\\
    \beta v_{pc}(r)=\dfrac{Z_pZ_cL_B}{\sqrt{r^2+(r_p+h)^2}}e^{-\sqrt{r^2+(r_p+h)^2}/\lambda},\\
    \beta v_{cc}(r)=\dfrac{Z_c^2L_B}{r}e^{-r/\lambda},
\end{array} \right.
\end{equation}
where    the Bjerrum length   $L_B=\dfrac{\beta e^2}{4\pi
\varepsilon_0\varepsilon_s}$, $\lambda$
 is the screening length of the solution, and $\beta=1/k_BT$. $Z_c$
and $Z_p$ are the effective charges of charged lipids and
macroions, respectively. Substituting Eq.\eqref{eq:model} into
Eqs.(1)-(3) gives the pair potentials we adopt here. Combined with
the hard-core repulsion of macroions, we can study the effects  of
the screening length, charges, and macroion's size, etc.

The three-component OZ equations are calculated in the k-space.
With some algebra, it can be written in a form more suitable for
numerical calculations, which is formally given by
\begin{equation}\label{3oz}
\tilde{h}_{\alpha\beta}(k)=f_{\alpha\beta}(\textbf{$\sigma$},\tilde{c}(k)),
\end{equation}
where $\tilde{c}(k)$=
$(\tilde{c}_{cn}(k),\tilde{c}_{np}(k),\tilde{c}_{pc}(k),\tilde{c}_{cc}(k),\tilde{c}_{nn}(k),
\tilde{c}_{pp}(k))$ and $\sigma$= $(\sigma_c,\sigma_n,\sigma_p)$. In
the limiting $\sigma_p\rightarrow0$, the OZ equations in the k-space
reduce to a two-component OZ equations of lipids:
\begin{subequations}\label{2oz}
\begin{align}
\tilde{h}_{cc}(k)&=\frac{\tilde{c}_{cc}(k)+\sigma_n(\tilde{c}_{cn}^2(k)-\tilde{c}_{cc}(k)\tilde{c}_{nn}(k))}{\tilde{D}(k)},\\
\tilde{h}_{nn}(k)&=\frac{\tilde{c}_{nn}(k)+\sigma_c(\tilde{c}_{cn}^2(k)-\tilde{c}_{cc}(k)\tilde{c}_{nn}(k))}{\tilde{D}(k)},\\
\tilde{h}_{cn}(k)&=\frac{\tilde{c}_{cn}(k)}{\tilde{D}(k)},
\end{align}
\end{subequations}
and   coupling equations with  macroions:
\begin{subequations}\label{couple}
\begin{align}
\tilde{h}_{pc}(k)&=\frac{\tilde{c}_{pc}(k)(1-\sigma_n\tilde{c}_{nn}(k))+\sigma_n\tilde{c}_{cn}\tilde{c}_{pn}(k)}{\tilde{D}(k)},\\
\tilde{h}_{pn}(k)&=\frac{\tilde{c}_{pn}(k)(1-\sigma_c\tilde{c}_{cc}(k))+\sigma_c\tilde{c}_{cn}\tilde{c}_{pc}(k)}{\tilde{D}(k)},
\end{align}
\end{subequations}
where
$\tilde{D}(k)=(1-\sigma_c\tilde{c}_{cc}(k))(1-\sigma_{n}\tilde{c}_{nn}(k))-\sigma_c\sigma_n\tilde{c}_{cn}^2(k)$.

The two-dimensional Fourier transformation of the isotropic system
can be  defined by
\begin{subequations} \label{fourier}
\
\begin{align}
     f(r)=\frac{1}{2\pi}\int_0^\infty k\tilde{f}(k)J_0(kr)\mathrm{d}k,\\
    \tilde{f}(k)=2\pi\int_0^\infty rf(r)J_0(kr) \mathrm{d}r,
\end{align}
\end{subequations}
where $J_0(x)$ is the zeroth-order Bessel function of the first
kind. In numerical calculations, Eq.\eqref{fourier} is truncated
and discretized in $\it r$-space and $\it k$-space. The discrete
Fourier transformation adopt here is a slow method, but satisfies
the orthogonality conditions (34,35) which are important in our
calculation.

The OZ equations  \eqref{2oz} are closed by the well-known PY
approximation, which is good for short-range potential and predicts
with a reasonable accuracy both structural and thermodynamic
properties of the investigated system (25). For the short-range
repulsion between the lipids studied here, it is obvious that the PY
approximation is more suitable. On the other hand, macroion's effect
can effectively be treated as an external field, and thus
Eq.\eqref{couple} is solved by combining with HNC approximation
(26), which is better for the long-range potential.
 Finally, the integral equations are solved by using the Picard
iteration scheme, and the  equality  \eqref{bridge} can be
rewritten as follows:
\begin{equation}
c_{\alpha\beta}^{new}(r)=p[e^{-\beta
V_{\alpha\beta}(r)+\gamma_{\alpha\beta}(r)+B_{\alpha\beta}(r)}-1-\gamma_{\alpha\beta}(r)]
+(1-p)c_{\alpha\beta}^{old}(r),
\end{equation}
where $p$ is an adjustable parameter to achieve the best
performance of   iterations.

\section{Results and Discussion}
In the infinite-diluted case, the coupling between  macroions has
been ignored, and thus the distribution function of macroions can
effectively be defined by $
 g_{pp}(r)=e^{-\beta V_{pp}^{eff}(r)}$. Because of the coupling between the macroions and the lipids, the
explicit distribution function  is written as follows:
\begin{equation}
g_{pp}(r)=e^{-\beta V_{pp}(r)+\sigma_c \int
h_{pc}(\textbf{r}')c_{pc}(|\textbf{r-r}'|)\mathrm{d}\textbf{r}'+\sigma_n
\int
h_{pn}(\textbf{r}')c_{pn}(|\textbf{r-r}'|)\mathrm{d}\textbf{r}'},
\end{equation}
namely the effective pair potential between  macroions is
\begin{equation}
\beta V_{pp}^{eff}(r)=\beta V_{pp}(r) -\sigma_c \int
h_{pc}(\textbf{r}')c_{pc}(|\textbf{r-r}'|)\mathrm{d}\textbf{r}'-\sigma_n
\int
h_{pn}(\textbf{r}')c_{pn}(|\textbf{r-r}'|)\mathrm{d}\textbf{r}'.
\end{equation}
Here, the effects of the lipids on the macroions have been
included in the effective pair potential between the macroions.

In the calculations presented below, we will restrict ourselves to
the case that different species of lipids have the same size,
$r_{ch}=r_{nh}=\tau$, $r_{cs}=r_{ns}=4\tau/3$. The macroion's
radius $r_p=4\tau$ and the minimal membrane-macroion distance
$h=0.3\tau$. The Fourier transformation is truncated at
$r\approx90\tau$. The choice of the soft-core range seems a bit
arbitrary, however, since such a range is fixed during our
calculation, it will not obscure the results given below. To keep
in line with the data appeared in Ref. (12), we choose
$r_{ch}=r_{nh}=\tau=4\AA$.

In Fig.~2 we show the effective potentials $\beta V_{pp}^{eff}(r)$
and the membrane-mediated potential $\beta V_{pp}^m(r)=\beta
V_{pp}^{eff}(r)-\beta V_{pp}(r)$ for varying values of the
demixing strength $\beta\omega_{cn}$ between neutral and charged
lipids in the physiological condition ($\lambda=10\AA,T=300K$).
The lipid concentration is fixed to be $\sigma_c\tau^2=0.055$ for
charged lipids and $\sigma_n\tau^2=0.105$ for neutral lipids. We
see that with increasing the repulsion between the neutral and the
charged lipids, the attraction between   macroions is greatly
enhanced. The result supports the conclusion made by Hinderliter
(7,8) from experiments and Monte-Carlo simulations. Obviously the
effective potential is composed of two parts: one is from the
contribution of the screened potential between macroions, and the
other is mediated by the substrate membrane. The membrane-mediated
attraction between macroions   is within $4 \sim 5$ lipid's
diameter from their surfaces.

Figure 3 shows  the total correlation functions $h_{pc}$ and
$h_{pn}$ with the lateral separation between macroion and lipid
for different values of $\beta\omega_{cn}$. The correlations
increases with the increase of $\beta\omega_{cn}$. Therefore, we
can say that the attraction is mediated by the membrane through
the correlations between the macroions and lipids. The resulting
local density distribution $\sigma_\alpha(r)$ of the $\alpha$
lipids which deviates from the averaged bulk density
$\sigma_\alpha$, can be obtained by the relation
$\sigma_\alpha(r)=(h_{p\alpha}(r)+1)\sigma_\alpha$, when a
macroion is in the center ($r=0$).  Interestingly, the density
distribution of the lipids under the macroions  is greatly
modified by increasing  the unmixing interaction
$\beta\omega_{cn}$ between neutral and charged lipids, while the
electrostatic interactions between the macroions and   charged
lipids keep unchanged. As $\beta\omega_{cn}=0$, the solid line in
Fig. 3 shows that the  enrichment of
 charged lipids  appears below the macroion, just through the screened Coulomb potential.
 Such a process can be  balanced by the reduction of the mixing
entropy and the increase of the repulsions between charged lipids.
As $\beta\omega_{cn}$ turns on, it has a tendency to drive the
same lipids  to aggregate together. This will greatly increase the
macroion-lipid correlations, as shown in Fig.~3, by enhancing the
aggregation of the same lipids, due to a pre-enrichment of charged
lipids below the macroion at $\beta\omega_{cn}=0$.

 The density profiles obtained here is similar to the
results obtained from   numerical solutions of the
Poisson-Boltzmann equation (12,13). However in the present case,
we have taken into account the correlation effects of species, and
from the inset of Fig.~3(b), we find that after an enriched region
of charged lipids, there is a depleted region of charged lipids
where the number density of charged lipids is even smaller than
the bulk case. Such a depleted effect of charged lipids will
become stronger if the correlation effect is enhanced, as shown in
Fig. 4 when $Z_c$ is increased to 2e. This is in agreement with
the statement about the multivalent system where  the charge
correlations become important, and thus there exists some
counter-intuitive effects (16), which are beyond the descriptions
of the PB theory.

For inhomogeneous mixing lipid system, if the interface between
neutral and charged lipids is absolutely sharp on a molecular
scale, the line tension will be directly proportional to the
strength of unmixing $\beta \omega _{cn}$. Therefore, to some
extent,  Figs. 2 and 3 indicate  that such an attractive effect
can be enhanced by the line tension between different lipids.
However, the weak attraction exists, even for the case
$\beta\omega_{cn}= 0$ where the line tension contributes little to
the attraction. It indicates a deeper screening of the
interactions between binding macroions in the presence of charged
lipids. The origin of such a potential is due to the bridging
effect of charged lipids, namely two macroions jointly attract to
the same set of charged lipids (28). Intuitively, it seems that
there exists another effect, the so-called depletion attractive
interaction (36), to contribute to $\beta V_{pp}^m(r)$. If we take
the charged lipids-rich clusters under the binding macroions as
static, it is possible to introduce such a depletion effect by
hard-core repulsions between the clusters and the surrounding
lipids. However, such a picture is naive. The
  charged lipids-rich regime is not static, but quite dynamic. The
charged and neutral lipids can move in and out   of the regime
quiet freely, although the density profiles remain  unchanged.
Therefore, there is no static interactions between the charged
lipids-rich regimes and the surrounding lipids. Such a dynamic
effect is similar to that of  Karanikas and Louis (28). To
illustrate it clearly, we eliminate the bridging effect by
introducing a pseudo potential between the binding macroions and
the charged lipids:
\begin{equation} \label{eq:psuedo}
 \left\{ \begin{array}{ll}
\beta V_{pc}(r)=\dfrac{Z_pZ_cL_B}{\sqrt{r^2+(r_p+h)^2}}e^{-\sqrt{r^2+(r_p+h)^2}/\lambda}, &r\leq 2r_p,\\
    \beta V_{pc}(r)=0, &r\geq 2r_p.
\end{array} \right.
\end{equation}
Such a potential does not allow two macroions to attract the same
set of  charged lipids, but the   charged lipids-rich regime still
exists. When we apply such a potential to our model in the case
$\beta \omega_{cn}=0$,   the weak attraction disappears. This
means that the bridging effect is dominant and there is no
depletion effect under the present dynamic picture.

In Fig.~5, we  change the relative composition of neutral and
charged lipids, but  the total number density of  lipids is fixed.
Interestingly, when $\sigma_c$ is low, the effective potential is
mainly repulsive, where the screened potential dominates over the
interactions between  macroions. With the increase of  charged
lipids, the membrane-mediated attraction becomes significant, and
it competes with the screened potential at a distance about $2
\tau$ - $6\tau$ and then forms an attractive well at that range.
The attractive force saturates when $\sigma_c\tau^2=0.05$. Further
increase of charged lipids will weaken the attractive potential,
because the screened potential  begins to dominate the system
again.

To clarify  the dominant contribution to the attractive potential,
we can calculate    the line tension by  varying the number
density of charged lipids, but   $\beta \omega_{cn}$ is fixed. In
the continuum limiting, the line tension is given by
$\gamma_l=C(\omega_{cn})\int\mathrm{d}\textbf{r}(\nabla\eta_{c}(r))^2$,
where  $\eta_{c}(r)$ is the local composition of
 charged lipids, and is estimated by
$\eta_c(r)=\sigma_c(r)/(\sigma_c(r)+\sigma_n(r))$. $C(\omega_{cn})$
is a coefficient relevant to $\omega_{cn}$, here it is a constant,
and thus we will present the result of $\gamma_l(r)$ in units of
$C(\omega_{cn})$. In Fig.~6, the solid square line tension curve
$\gamma_l(r)/C(\omega_{cn})$   is plotted as a function of  the
number density of charged lipids for the infinite-diluted case
($\sigma_p \tau^2\rightarrow0$). The line tension first increases
with $\sigma_c \tau^2$, and then is decreased.  The maximum line
tension
 almost correspond to  the deepest attractive potential of
  lipid-mediated interactions between macroions in Fig.5,
 and they are of the same order. We can conclude that the  attractive potential increases
with  the line tension,   depending upon two factors: the demixing
interaction and the relative component  between neutral and
charged lipids. On the other hand, the effective attractive
potential between macroions is always enhanced by increasing
charged lipid's concentration $\sigma_c\tau^2$, namely the
attraction due to the bridge effect monotonously increases with
$\sigma_c\tau^2$. Therefore,  there is a cooperation between the
line tension and the bridging effect which jointly  assists the
attraction between macroions at the beginning stage of the
increase of $\sigma_c\tau^2$, and for relatively large values of
$\sigma_c\tau^2$, a competition occurs, since line tension will
decrease with further increase of charged lipids.  The fact that
the line tension behavior in Fig.6 is in accordance with the
effective potential well vs $\sigma_c \tau^2$ in Fig.5, clearly
indicates that the line tension plays a vital important role in
the membrane-mediated potentials, as argued in the Ref. (12) in
the continuum limits of the membrane.

  We now turn to discuss the effects of finite macroion
concentration. In this case, the coupling between   macroions
cannot be ignored, and we have to solve the three-component OZ
equations. The effective potential $V^{eff}_{pp}$ between
macroions can be obtained by an inverse process (37). After
solving the integral equation Eq.(7), we can obtain the total
correlation function $h_{pp}(r)$, and then the direct correlation
function $\tilde{c}_{eff}(k)$ can be solved from the reduced
one-component OZ equation:
\begin{equation}
\tilde{c}^{eff}(k)=\dfrac{\tilde{h}_{pp}(k)}{1+\sigma_p\tilde{h}_{pp}(k)}\;.
\end{equation}
The effective potential can be solved from the HNC approximation
(24,37):
\begin{equation}\label{inversion}
g_{pp}(r)=e^{-\beta V_{pp}^{eff}(r)+h_{pp}(r)-c^{eff}(r)}.
\end{equation}
As $\sigma_{p}\rightarrow 0$,
 the present calculation reduces to
the infinite-diluted case.

In Fig.~6, we firstly show the line tension as a function of
$\sigma_{c}\tau^2$ for  different macroion concentrations. We
notice that on increasing binding macroions, the line tension is
gradually decreased and the position of the maximum line tension
shifts toward the higher values of $\sigma_{c}$. These two
behaviors are of the same origin, because there exists a competing
interplay of different macroions to recruit charged lipids. The
competition will partially suppress the modification of lipid
composition profiles induced by a single macroion, and thus the
line tension decreases, in comparison with the case $\sigma_p
\tau^2\rightarrow0$. Furthermore, we can expected that, at lower
macroion concentrations, the charged lipid's profile saturates
under a single macroion and the line tension reaches the maximum
at a certain value of $\sigma_c \tau^2$. However, if the macroion
concentration is increased, the charged lipid's concentration will
not   saturate at a former value of $\sigma_c \tau^2$. To achieve
the maximum line tension, we should further increase the charged
lipid's concentration, i.e., the maximum line tension occurs with
a higher value of  $\sigma_c \tau^2$ with the increase of  binding
macroions.

 In Fig.7 we depict the effective potential for varying values of $\sigma_c \tau^2$  when $\sigma_p
 \tau^2=0.004$.  We can see that the change of  effective potentials has the
same tendency as  in the infinite-diluted case. When the
concentration of charged lipids is low, the screened Coulomb
potential dominates, and the effective potential is repulsive. In
an intermediate concentration, the membrane-mediated attractive
potentials compete with the screened Coulomb potentials and form
an effective attractive well. With further increase of  charged
lipids, the mediated attractive potentials decay, and the screened
Coulomb potentials dominate over the system again. If we calculate
the line tension in this case, also, it has the same tendency as
in the infinite-diluted case. However, we see from Fig.7 that  on
increasing the binding macroions, the attractive potential
saturates at a certain concentration with $\sigma_c\tau^2=0.08$,
which deviates slightly from the value $\sigma_c\tau^2=0.07$ where
the line tension is maximum (Fig.~6). Such a deviation indicates
that although line tension is still dominant for finite
concentration of macroions, the bridge effect will become
relatively stronger, compared to the infinite-diluted case.

\section{Conclusion}

In this paper, we have studied the membrane-mediated potential
between the binding macroions which is induced by the interplay of
the macroions and lipids. The process can be generally interpreted
as follows.
 When the binding macroions are infinite dilute, the
macroion will induce an enrichment of charged lipids under it.
This process will be balanced by the decrease of the mixing
entropy and the increase of repulsion between charged lipids. When
there is a soft-core repulsion between  charged and   neutral
lipids, the nonuniform distribution of   lipids will introduce a
line tension. The macroions prefer to aggregate together to reduce
the line tension energy. This will compete with the screened
Coulomb potential and form an attractive well in the effective
potential. The bridging effect will also contribute to the
attractive potential, but the line tension is dominant.
Interestingly, as we increase the number density of  charged
lipids, but fix the total number density of   lipids, we observe a
repulsion - attraction - repulsion transition due to the
competition between the line tension and screened electrostatic
interactions. For the finite concentration case, we have to take
into account  the coupling effects between binding macroions. We
find that with the increase of   binding macroions, the line
tension induced by a single binding macroion is gradually
suppressed and the maximum of attractive potential will shift
toward the higher values of charged lipid concentration
$\sigma_c$. The result can be attributed to the competition of
recruiting charged lipids due to different binding macroions. The
effective potential calculated in this case is not as faithful as
the infinite dilution case. But in the case where the
concentration of binding macroions is fixed, the result obtained
is physically reasonable.

This work was supported by  the National Natural Science
Foundation of China, No. 10334020, No. 10021001,  No. 20490220,
and No. 10574061.

\newpage

Figure captions:

 Figure~1: Schematic of a macroion bound to a binary fluid lipid membrane. The lipid is
treated as a hard-sphere head with a radius of $r_{ch}$(charged
lipid) or $r_{nh}$(neutral lipid) and a soft-core repulsion tails.
The binding macroions are treated as uniformly charged hard
spheres with radius $r_{p}$. There is a separation $h$ from the
membrane to the bottom of the macroion. The system is a
three-dimensional system, but  can be effectively mapped into a
two-dimensional one when $h$ is fixed.

Figure~2: The effective pair potentials $\beta V_{pp}^{eff}(r)$ or
$\beta V_{pp}^m(r)$ ($=\beta V_{pp}^{eff}(r)-\beta V_{pp}(r)$)
change with $\beta\omega_{cn}$. The macroion's diameter is chosen
to be 8$\tau$, and the effective charges of the macroions and
charged lipids is $Z_p=6e$ and $Z_c=-1e$, respectively. The
mediated potential $\beta V_{pp}^m(r)=\beta V_{pp}^{eff}(r)-\beta
V_{pp}(r)$ becomes more attractive with the increase of
$\beta\omega_{cn}$. This is a general behavior of the system.

Figure~3: The total correlation functions $h_{pn}(r)$ and
$h_{pc}(r)$ between the lipids and macroion for varying values of
$\beta\omega_{cn}$.  In (b), the curves for the lateral distance
$r
> 12\tau$ is enlarged, and re-plotted in the inset, showing the
correlation effects which will not appear in the mean field
approximation.

Figure~4: The total correlation function $h_{pc}(r)$ verse
$\beta\omega_{cn}$ for $Z_c=-2e$.

Figure~5: Effective pair potential $\beta V_{pp}^{eff}(r)$ verse the
number density of   charged lipids for fixed  total lipids
concentration   $(\sigma_c+\sigma_n)\tau^2=0.16$ and
$\beta\omega_{cn}=2.0$. The attractive potential saturates at an
intermediate concentration ($\sigma_c\tau^2=0.05$) of charged
lipids.

Figure~6: Single macroion induced line tension as a function of
$\sigma_c\tau^2$ for fixed  total lipids concentration
$(\sigma_c+\sigma_n)\tau^2=0.16$ and $\beta\omega_{cn}=2.0$. The
solid square curve shows the case $\sigma_p\tau^2\rightarrow0$,
and the line tension saturates at $\sigma_c\tau^2=0.05$. As the
binding macroions are increased, The line tension decreases, and
the position of maximum line tension    moves toward the higher
values of $\sigma_c \tau^2$.

Figure~7: Effective pair potential $\beta V_{pp}(r)$ for varying
concentrations  of charged lipids with
$(\sigma_c+\sigma_n)\tau^2=0.16$,  $\beta\omega_{cn}=2.0$,  and
$\sigma_p\tau^2=0.004$. The attractive potential saturates at an
intermediate concentration $\sigma_c\tau^2=0.08$.

\newpage

\begin{figure} \label{fig:dud}
\includegraphics[width=17.8cm]{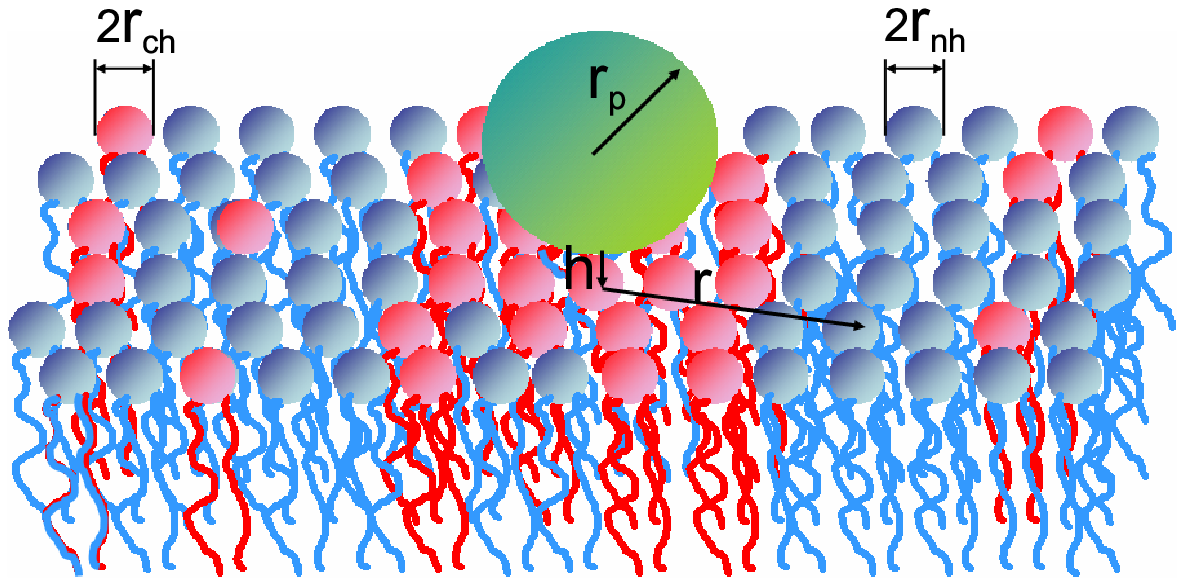}
\caption{}
\end{figure}

\begin{figure} \label{fig:bd}
\epsfxsize=17.8cm  \epsfbox{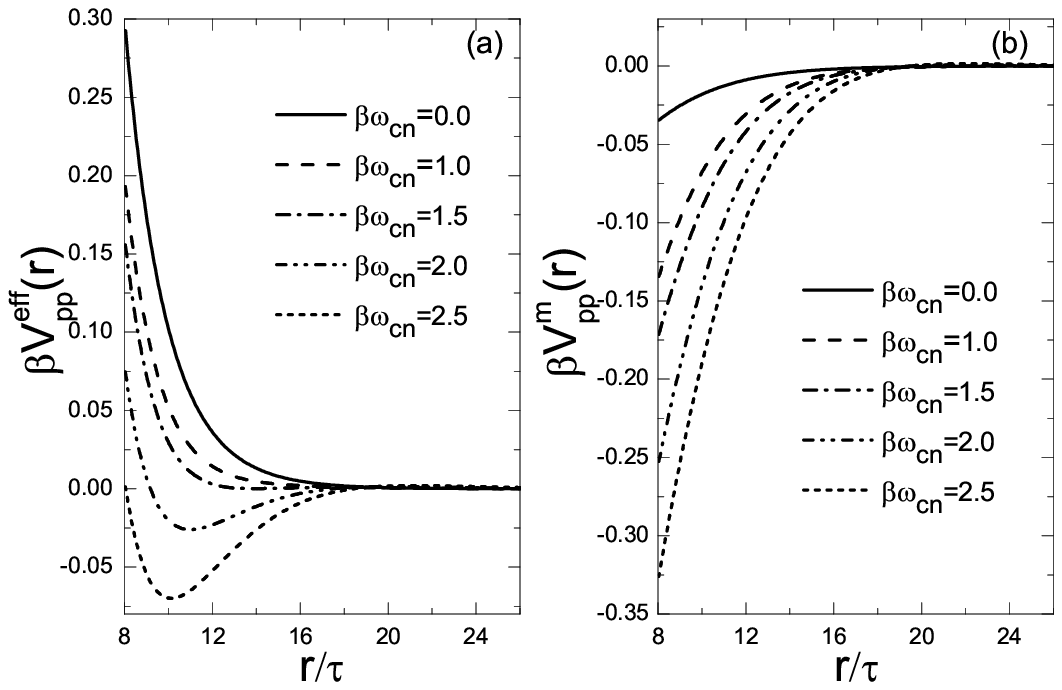}\caption{}
\end{figure}

\begin{figure} \label{fig:cd}
\epsfxsize=17.8cm  \epsfbox{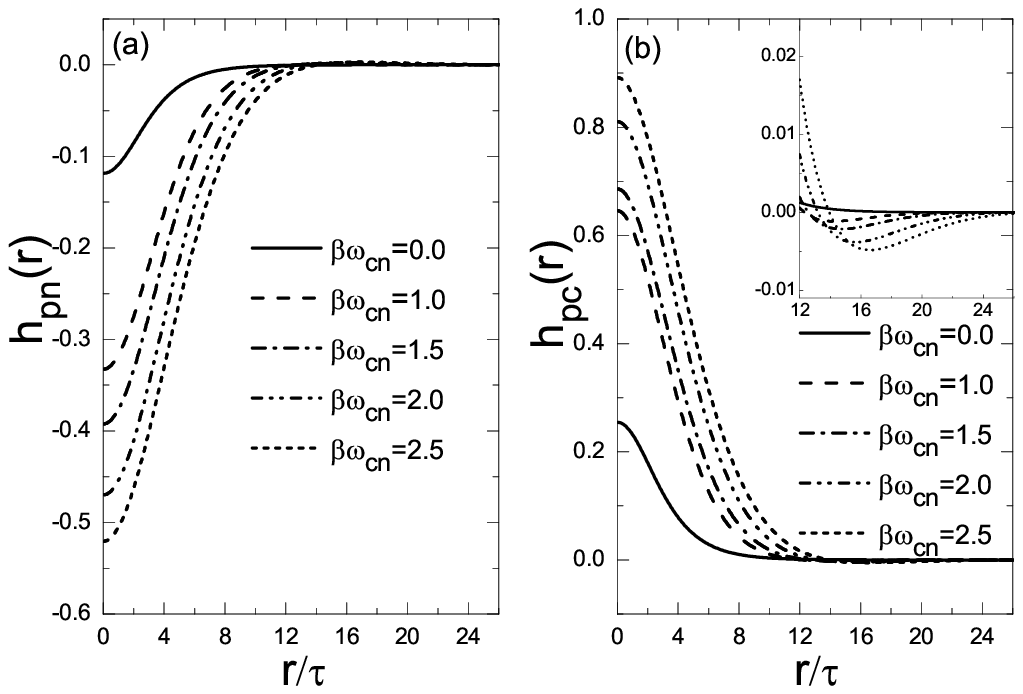}\caption{ }
\end{figure}

\begin{figure} \label{fig:cd}
\epsfxsize=17.8cm  \epsfbox{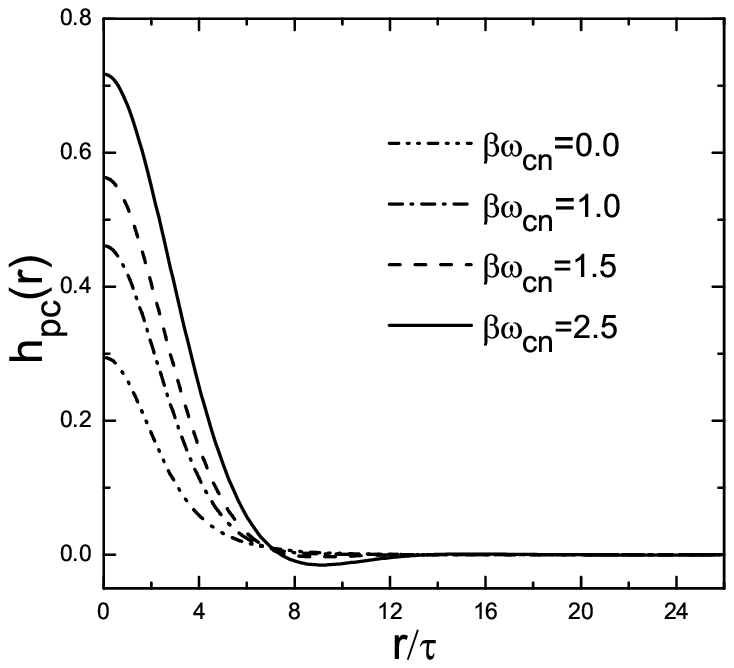}\caption{}
\end{figure}

\begin{figure} \label{fig:44cd}
\includegraphics[width=17.8cm]{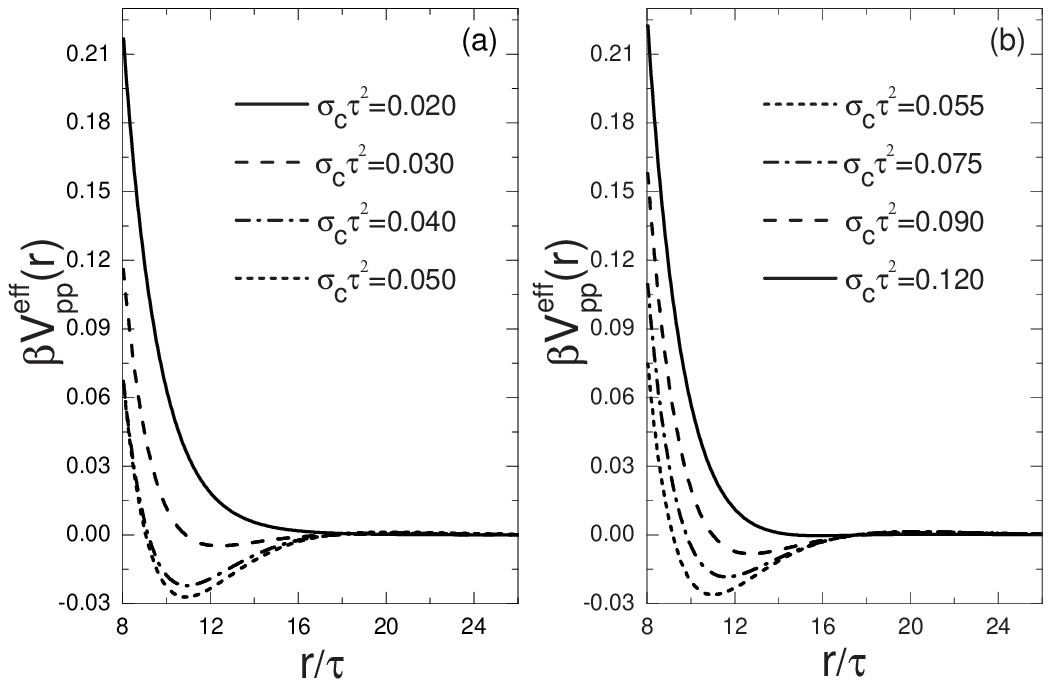}
\caption{}
\end{figure}

\begin{figure} \label{fig:dbd}
\includegraphics[width=17.8cm]{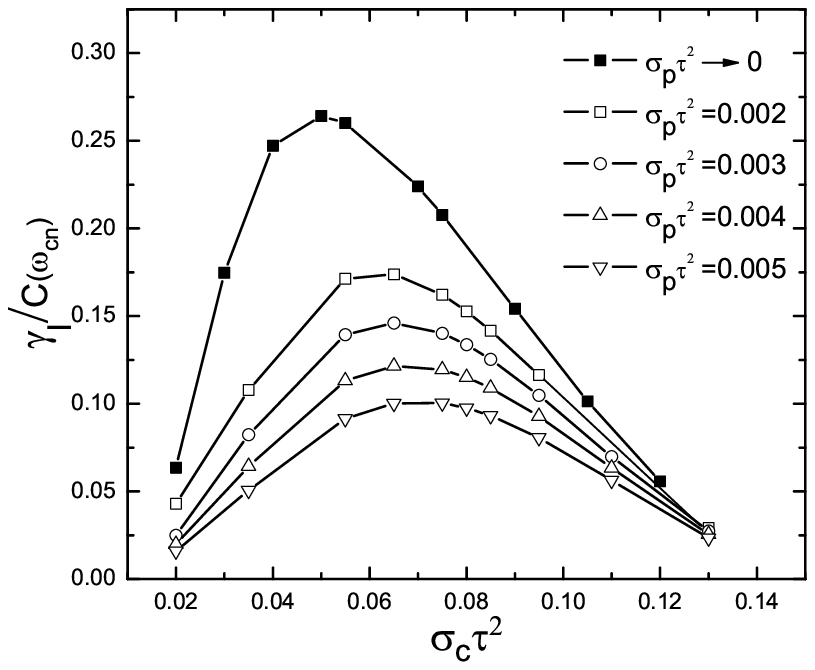}
\caption{}
\end{figure}

\begin{figure} \label{fig:dujd}
\includegraphics[width=17.8cm]{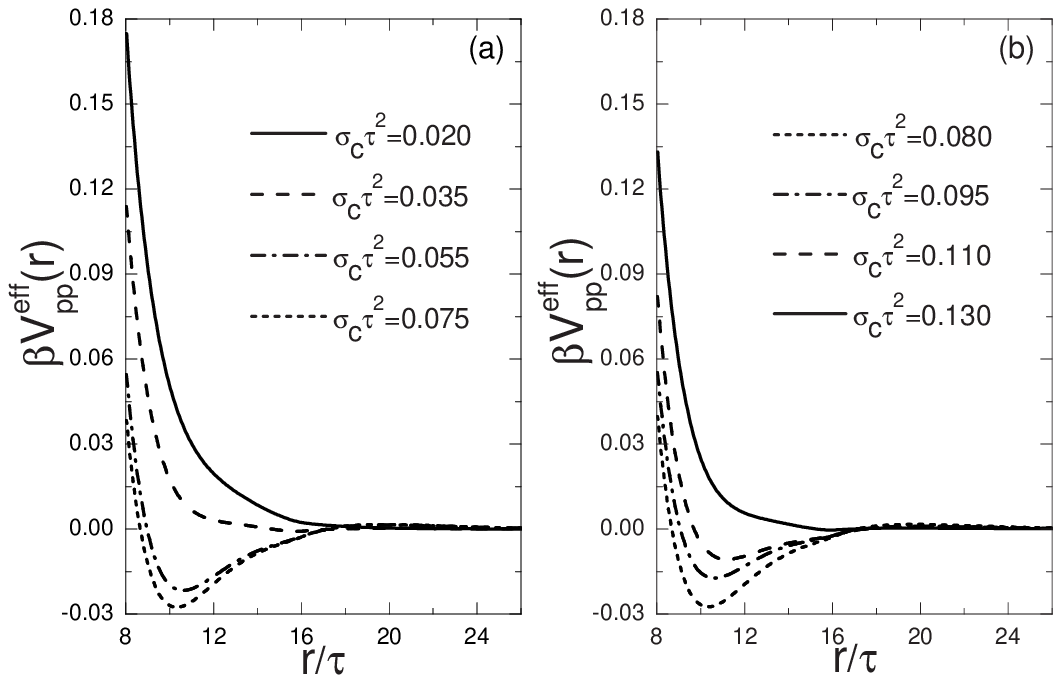}
\caption{}
\end{figure}


\begin{references}

1. Bongrand, P. 1999. Ligand-receptor interactions. \textit{Rep.
Prog. Phys.} 62:921-968.\\

2. Simons, K., and E. Ikonen. 1997. Functional rafts in cell
membranes. \textit{Nature.} 387:569-572.\\

\noindent3. Lipowsky, R., and R. Dimova. 2003. Domains in
membranes and vesicles. \textit{J. Phys.: Condens. Matter.}
15:S31-S45.\\

\noindent 4. Ben-Tal, N., B. Honig, R. M. Peitzsch, G. Denisov,
and S. Mclaughlin. 1996. Binding of small basic peptides to
membranes containing acidic lipids: theoretical models and
experimental results. \textit{Biophys. J.} 71:561-575.\\

\noindent 5. Denisov, G., S. Wanaski, P. Luan, M. Glaser, and S.
Mclaughlin. 1998. Binding of basic peptides to membranes produces
lateral domains enriched in the acidic lipids phosphatidylserine
and phosphatidylinositol 4,5-bisphosphate: an electrostatic model
and experimental results. \textit{Biophys. J.} 74:731-744.\\

\noindent 6. Heimburg, T., B. Angerstein, and D. Marsh. 1999.
Binding of peripheral proteins to mixed lipid membranes: effect of
lipid demixing upon binding. \textit{Biophys. J.} 76:2575-2586.\\

\noindent 7. Hinderliter, A., P. F. F. Almeida, C. E. Creutz, and
R. L. Biltonen. 2001. Domain Formation in a fluid mixed lipid
bilayer modulated through binding of
    the C2 protein motif. \textit{Biochemistry.}
40:4181-4191.\\

\noindent 8. Hinderliter, A., R. L. Biltonen, and P. F. F. Almeida.
2004. Lipid modulation of protein-induced membrane domains as a
mechanism for
    controlling signal transduction. \textit{Biochemistry.}
    43:7102-7110.\\

\noindent 9. May, S., D. Harries, and A. Ben-Shaul. 2000. Lipid
demixing and protein-protein interactions in the adsorption of
    charged proteins on mixed membranes. \textit{Biophys. J.}
    79:1747-1760.\\

\noindent 10. May, S., D. Harreis, and A. Ben-Shaul. 2002.
Macroion-induced compositional instability of binary fluid
membranes. \textit{Phys. Rev. Lett.} 26:268102.\\

\noindent 11. May, S. 2005. Stability of macroion-decorated lipid
membranes. \textit{J. Phys.: Condens. Matter.}
 17:R833-R850.\\

\noindent 12. Mbamala, E. C., A. Ben-Shaul, and S. May. 2005.
Domain formation induced by the adsorption of charged proteins on
mixed lipid membranes. \textit{Biophys. J.} 88:1702-1714.\\

\noindent 13. Haleva, E., N. Ben-Tal, and H. Diamant. 2004.
Increased concentration of polyvalent phospholipids
 in the adsorption domain of a charged protein. \textit{Biophys. J.}
 86:2165-2178.\\

\noindent 14. Neu, J. C. 1999. Wall-mediated forces between
like-charged bodies in an electrolyte. \textit{Phys. Rev. Lett.}
82:1072-1074.\\

\noindent 15. Naji, A., S. Jungblut, A. G. Moreira, and R. R.
Netz. 2005. Electrostatic interactions in strongly coupled soft
matter. \textit{Physica A.} 352:131-170.\\

\noindent 16. Levin, Y. 2005. Strange electrostatics in physics,
chemistry, and biology.
 \textit{Physica A.} 352:43-52.\\

\noindent 17. Levin, Y. 2002. Electrostatic correlations: from
plasma to biology. \textit{Rep. Prog. Phys.} 65:1577-1632.\\

\noindent 18. Wu, J., D. Bratko, and J. M. Prausnitz. 1998.
Interaction between like-charged colloidal spheres in electrolyte
solutions. \textit{Proc. Natl. Acad. Sci. USA.} 95:15169-15172.\\

\noindent 19. Allahyarov, E., I. D'Amico, and H. L\"{o}wen. 1998.
Attraction between like-charged macroions by coulomb depletion.
\textit{Phys. Rev. Lett.} 81:1334-1337.\\

\noindent 20. Linse, P., and  V. Lobaskin. 1999. Electrostatic
attraction
 and phase separation in solutions of like-charged colloidal particles. \textit{Phys. Rev.
Lett.} 83:4208-4211.\\

\noindent 21. Hribar, B., and V. Vlachy. 2000. Clustering of
macroions in solutions
 of highly asymmetric electrolytes. \textit{Biophys. J.}
 78:694-698.\\

\noindent 22. Kepler, G. M., and S. Fraden. 1994. Attractive
potential between confined
 colloids at low ionic strength. \textit{Phys. Rev. Lett.}
 73:356-359.\\

\noindent 23. Behrens, S. H., and D. G. Grier. 2001. Pair
interaction of charged
 colloidal spheres near a charged wall. \textit{Phys. Rev. E.}
 64:050401.\\

\noindent 24. Han, Y. L., and D. G. Grier. 2003.
Confinement-induced colloidal attractions
 in Equilibrium. \textit{Phys. Rev. Lett.} 91:038302.\\

\noindent 25. Caccamo, C. 1996. Integral equation theory description
of phase equilibria in
 classical fluids. \textit{Phys. Rep.} 274:1-105.\\

\noindent 26. Hansen, J. P., and I. R. McDonald. 1986. Theory of
Simple Liquids, 2nd Ed. Academic Press, London.\\

\noindent 27. Tohver, V., J. E. Smay, A. Braem, P. V. Braun, and
J. A. Lewis. 2001. Nanoparticle halos: A new colloid stabilization
mechanism. \textit{Proc. Natl. Acad. Sci. USA.} 98:8950-8954.\\

\noindent 28. Karanikas, S., and A. A. Louis. 2004. Dynamic
colloidal stabilization by
 nanoparticle halos. \textit{Phys. Rev. Lett.} 93:248303.\\

\noindent 29. Chen, K., and Y. Q. Ma. 2005. Interactions between
colloidal particles induced by
 polymer brushes grafted onto the substrate. \textit{J. Phys. Chem. B.}
 109:17617-17622.\\

\noindent 30. Feller, S. E., R. M. Venable, and R. W. Pastor.
1997. Computer simulation of a DPPC phospholipid bilayer:
structural changes as a function of molecular surface area.
\textit{Langmuir.} 13:6555-6561.\\

\noindent 31. Lag\"{u}e, P., M. J. Zuckermann, and B. Roux. 2000.
Lipid-mediated interactions between intrinsic
 membrane proteins: a theoretical study based on integral equations. \textit{Biophys. J.}
79:2867-2879.\\

\noindent 32. Rosenfeld, Y., and N. W. Ashcroft. 1979. Theory of
simple classical fluids:
 universality in the short-range structure. \textit{Phys. Rev.
A.} 20:1208-1235.\\

\noindent 33. Coster, H. G. L. 2003. The physics of cell membranes
 \textit{J. Biol. Phys.} 29:363-399.\\

\noindent 34. Lado, F. 1971. Numerical fourier transforms in one,
two, and three dimensions
 for liquid state calculations.  \textit{J. Comput. Phys.}
 8:417-433.\\

\noindent 35. Yethiraj, A., B. J. Sung, and F. Lado. 2005. Integral
equation theory for
 two-dimensional polymer melts. \textit{J. Chem. Phys.}
 122:094910.\\

\noindent 36. Asakura, S., and F. Oosawa. 1954. On interaction
between two bodies immersed in a
 solution of macromolecules. \textit{J. Chem. Phys.}
 22:1255-1256.\\

\noindent 37. Reatto, L., D. Levesque, and J. J. Weis. 1986.
Iterative pedictor-corrector method for
 extraction of the pair interaction from structrual data for dense classical liquids. \textit{Phys. Rev.
A.} 33:3451-3465.\\
\end{references}
\end{document}